\documentclass [preprint,aps,showpacs] {revtex4}
\usepackage{bm}
\usepackage{epsfig}
\begin{document}

\title{Finite-height effect on electron energy structure of lead salts 
nanorods}

\author{S.V.~Goupalov}
\affiliation{Department of Physics, Jackson State University,
Jackson, MS 39217 USA\\
and A.F.~Ioffe Physico-Technical Institute, 26 Polytechnicheskaya, 194021 St.~Petersburg, Russia}

\begin{abstract}
The effect of the finite height of a cylindrical lead salt nanorod on its electronic structure
is studied within the effective mass formalism. It is demonstrated that for most practical purposes it can be accounted for by sampling the subband dispersion dependencies of the infinite cylindrical quantum wire of the same radius at wave numbers $k_z=\pi n_z/H$, where
$H$ is the nanorod height and $n_z$ is an integer. However, under certain conditions the otherwise degenerate electron energy levels will repel. A detailed account of these conditions is provided.
\end{abstract}

\pacs{73.21.Hb,73.22.Dj,78.67.Lt}
\maketitle

In recent years a significant progress has been made in solution-based synthesis
of semiconductor nanostructures. As a result, a new class of nanostructures called nanorods has emerged~\cite{alivisat}. The nanorods can be considered as an intermediate case between the quasi-zero-dimensional (0D) quantum dots and quasi-one-dimensional (1D) quantum wires and allow one to investigate the variation in material properties in the transition from 0D to 1D. They can be thought of as cylindrical 
semiconductor structures having diameter on the order of several nanometers and
height up to several tens of nanometers.

Recently nanorods of lead salt semiconductors have been synthesized by a number of research groups~\cite{acharya,luther,koh,rubin}.
Lead salt semiconductor compounds (PbS, PbSe, PbTe) are characterized by a large
bulk exciton Bohr radius, in the range of 20 to 46 nm. Thus, in lead salt
nanorods of a relatively short height, the strong confinement regime of charge
carriers can be achieved in all three dimensions. 

In Ref.~\cite{mycom} we have studied electronic structure of cylindrical
lead salt quantum wires within the effective mass approximation. We found that
the electron energy subbands of the conduction and valence bands in these quantum
wires have monotonous dispersion dependencies. At first glance, in this situation the effect
of finite dimension of the structure along the nanorod growth direction on the
electron energy structure could be accounted for by  
substitution of the discrete values of wave numbers, $\pi n_z/H$ (where $H$ is the
nanorod height and $n_z$ is an integer) into energy dispersion dependencies for various subbands of the quantum wire. It turns out, however, that such procedure
does not account for some qualitative changes in electron energy structure caused
by the finite height of the nanorod. In this paper we will first describe these
changes using a numerical solution of the effective mass equations, and then explain them based on the analytical dispersion equation derived in Ref.~\cite{mycom} for the case of a quantum wire.

The conduction and valence band extrema in lead salt semiconductors
occur at the $L$-points of the Brillouin zone leading to the high valley degeneracy. Bartnik {\it et al.} have studied solutions of the effective mass equations 
customized for each of the four $L$-valleys for different quantum wire growth
directions~\cite{bartnik}. They showed that the valleys remain degenerate, within
the effective mass approximation, when the growth occurs along the $\langle 100 \rangle$ direction. It is not
infrequent that lead salt nanorods grow along this crystallographic 
axis~\cite{koh,rubin,bartnik}. The valley effective masses in this case still remain anisotropic. In what follows we will neglect this anisotropy and assume that
the nanorods grow along the $\langle 100 \rangle$ direction. 
Then the electron spectrum near the $L$-point
taking into account only the two closely lying conduction
and valence bands can be described by the
spherical Dimmock model~\cite{mycom,kang}.
In this model the electron
wave function is written as
\begin{equation}
\label{Psi1}
\Psi= \hat{u} \, |L_6^- \rangle + \hat{v} \, |L_6^+ \rangle \,,
\end{equation}
where $|L_6^- \rangle$ and $|L_6^+ \rangle$ describe
the Bloch functions while $\hat{u}({\bf r})$ and $\hat{v}({\bf r})$
are the spinors slowly varying with coordinates and satisfying the
equation
\begin{equation}
\label{dimmock}
\hat{H}(-i {\bm \nabla}) \left[
\matrix{
\hat{u} \cr
\hat{v}
}
\right] =E \, \left[
\matrix{
\hat{u} \cr
\hat{v}
}
\right]
\end{equation}
with
\begin{equation}
\label{ham}
\hat{H}(-i {\bm \nabla})= 
\left[
\matrix{
\left( \frac{E_g}{2} - \alpha_c \, \Delta \right)
&
-i P \left( {\bm \sigma} {\bm \nabla} \right)
\cr
-i P \left( {\bm \sigma} {\bm \nabla} \right)
&
-\left( \frac{E_g}{2} - \alpha_v \, \Delta \right)
\cr
}
\right] \,
 \,.
\end{equation}
Here $\sigma_{\beta}$ ($\beta=x,y,z$) are the Pauli matrices, $\alpha_c$,
$\alpha_v$, $E_g$, and $P$ are parameters of the model and $E$ is the electron
energy. Eq.~(\ref{ham}) is written using the system of units where $\hbar=m_0=e=1$ ($m_0$ and
$e$ are the mass and the charge of a free electron). Inclusion of the valley 
anisotropy would amount to introduction of longitudinal and transverse counterparts
for the parameters $P$, $\alpha_c$, and $\alpha_v$~\cite{bartnik}.

In order to solve Eq.~(\ref{dimmock}) numerically for boundary conditions of cylindrical symmetry, one can first write
down the Hamiltonian~(\ref{ham}) in the basis of four cylindrically symmetric
bispinors of the form
\[
\Psi_1^{M,n}(\rho,\varphi,z)=\frac{e^{i(M-1/2) \varphi}}{\sqrt{2 \pi}} \, f(z) \, C_{M-1/2,n} \,
J_{M-1/2} \left( x_{M-1/2,n} \frac{\rho}{R} \right) \,
\left[
\matrix{
1 \cr
0 \cr
0 \cr
0}
\right] \,,
\]
\[ 
\Psi_2^{M,n}(\rho,\varphi,z)=\frac{e^{i(M+1/2) \varphi}}{\sqrt{2 \pi}} \, f(z) \, C_{M+1/2,n} \,
J_{M+1/2} \left( x_{M+1/2,n} \frac{\rho}{R} \right) \,
\left[
\matrix{
0 \cr
1 \cr
0 \cr
0}
\right] \,,
\] 
\[
\Psi_3^{M,n}(\rho,\varphi,z)=\frac{e^{i(M-1/2) \varphi}}{\sqrt{2 \pi}} \, f(z) \, C_{M-1/2,n} \,
J_{M-1/2} \left( x_{M-1/2,n} \frac{\rho}{R} \right) \,
\left[
\matrix{
0 \cr
0 \cr
1 \cr
0}
\right] \,,
\]
\[ 
\Psi_4^{M,n}(\rho,\varphi,z)=\frac{e^{i(M+1/2) \varphi}}{\sqrt{2 \pi}} \, f(z) \, C_{M+1/2,n} \,
J_{M+1/2} \left( x_{M+1/2,n} \frac{\rho}{R} \right) \,
\left[
\matrix{
0 \cr
0 \cr
0 \cr
1}
\right] \,.
\] 
Here $R$ is the radius of the cylindrical nanostructure,
$x_{L,n}$ is the $n$th zero of the Bessel function, $J_L(x)$, and
\[
C_{L,n}=\frac{\sqrt{2}}{R |J_{L+1}(x_{L,n})|}
\]
is the normalization coefficient. All these bispinors
are characterized by the quantum number $M$, the projection of the total angular
momentum onto the growth direction, $z$, and vanish on the cylindrical surface of
the nanostructure. 

For a cylindrical quantum wire the function $f(z)$ should be chosen in the form of a plane wave
\[
f(z)=\frac{e^{ik_zz}}{\sqrt{2 \pi}} \,.
\]
If the index $n$ runs from 1 to $n_{max}$, then the Hamiltonian~(\ref{ham}) written in the chosen basis, for given quantum numbers $M$ and $k_z$, represents a $4 \, n_{max} \times 4 \, n_{max}$ matrix. If one is interested
in a certain number of subbands characterized by the given value of the total angular momentum projection, $M$, it is enough to set $n_{max}$ equal to that
number. One can check that numerical diagonalization of the resulting Hamiltonian leads
to electron energy dispersion dependence coinciding (with a very high precision) 
with that obtained from solution of  
the dispersion equation derived analytically and given by Eq.~(25) of Ref.~\cite{mycom}. The numerical procedure is very robust and for practical calculations it would be our first choice. 

\begin{figure}[htb]
  \centering
    \includegraphics[width=.4\textwidth]{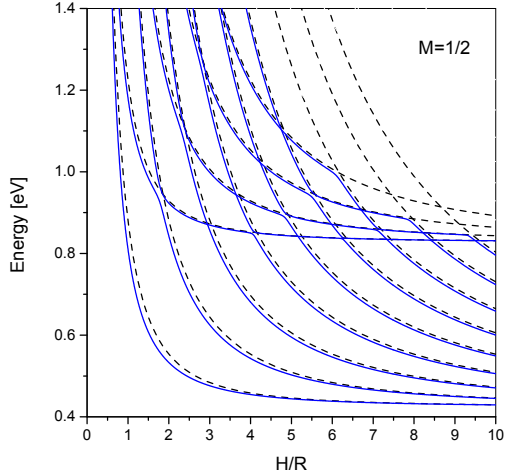}
\caption{(Color online) Energies of levels of size quantization described by the quantum number $M=1/2$ in a PbSe
nanorod of radius $R=20$~\AA\mbox{} and height $H$ as functions of $H/R$ (solid lines). Dashed lines: energy dispersion $E_{1/2,n_c}(k_z)$ for subbands 
of a PbSe cylindrical quantum wire of
radius $R=20$~\AA\mbox{} for quantum numbers $M=1/2$, $n_c=1,2$~\cite{mycom} calculated at $k_z=\pi n_z/H$.}
\end{figure}

For a nanorod of height $H$ the function $f(z)$ can be chosen in the form
\begin{equation}
f(z)=\sqrt{\frac{2}{H}} \sin{\frac{\pi n_z z}{H}} \,.
\label{frod}
\end{equation}
If the index $n_z$ runs from $1$ to $n_z^{max}$ then the Hamiltonian~(\ref{ham}),
for a given quantum number $M$, 
represents in the chosen basis a  
$4 \, n_{max} \, n_z^{max} \times 4 \, n_{max} \, n_z^{max}$ matrix. 
We have numerically diagonalized this matrix for nanorod radius $R=20$~\AA\mbox{} and 
nanorod heights in the range from $10$~\AA\mbox{} to $200$~\AA\mbox{}. We used the same 
material parameters as in Ref.~\cite{mycom} (corresponding to PbSe nanorods)
and the values of $n_{max}=3$, $n_z^{max}=30$. In Fig.~1 by solid lines are shown the dependencies 
of the first 10 eigenvalues occuring within the conduction band and corresponding
to the quantum number $M=1/2$ on the height to radius ratio, $H/R$. By dashed lines
in Fig.~1 are shown the quantum wire subband energies~\cite{mycom} $E_{1/2,n_c}(k_z)$ calculated at $k_z=\pi n_z/H$ for a PbSe quantum wire of radius $R=20$~\AA.

\begin{figure}[htb]
  \centering
    \includegraphics[width=.4\textwidth]{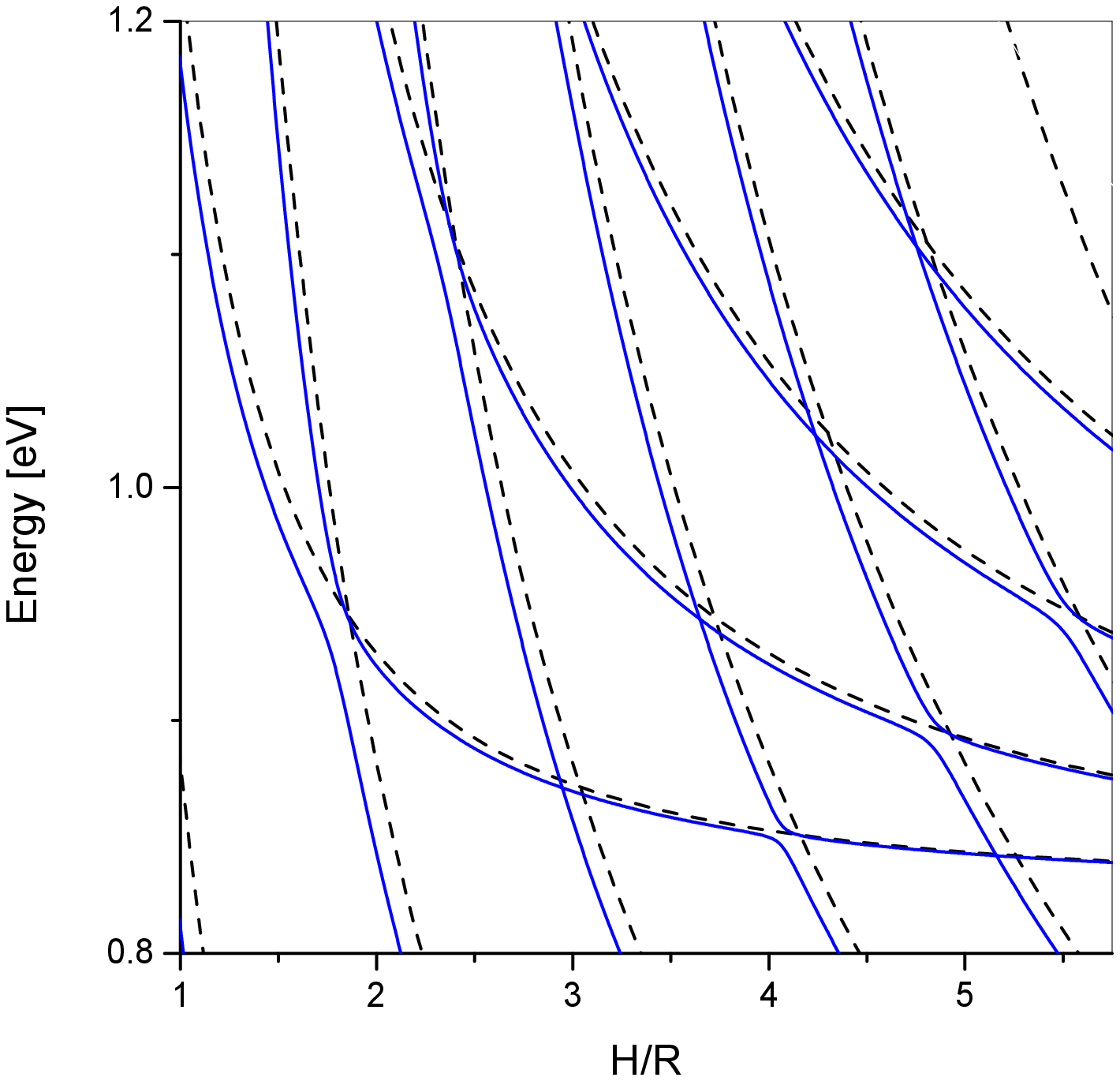}
\caption{(Color online) Same as Fig.~1 but on an
enlarged scale.}
\end{figure}
\begin{figure}[ht]
  \centering
    \includegraphics[width=.4\textwidth]{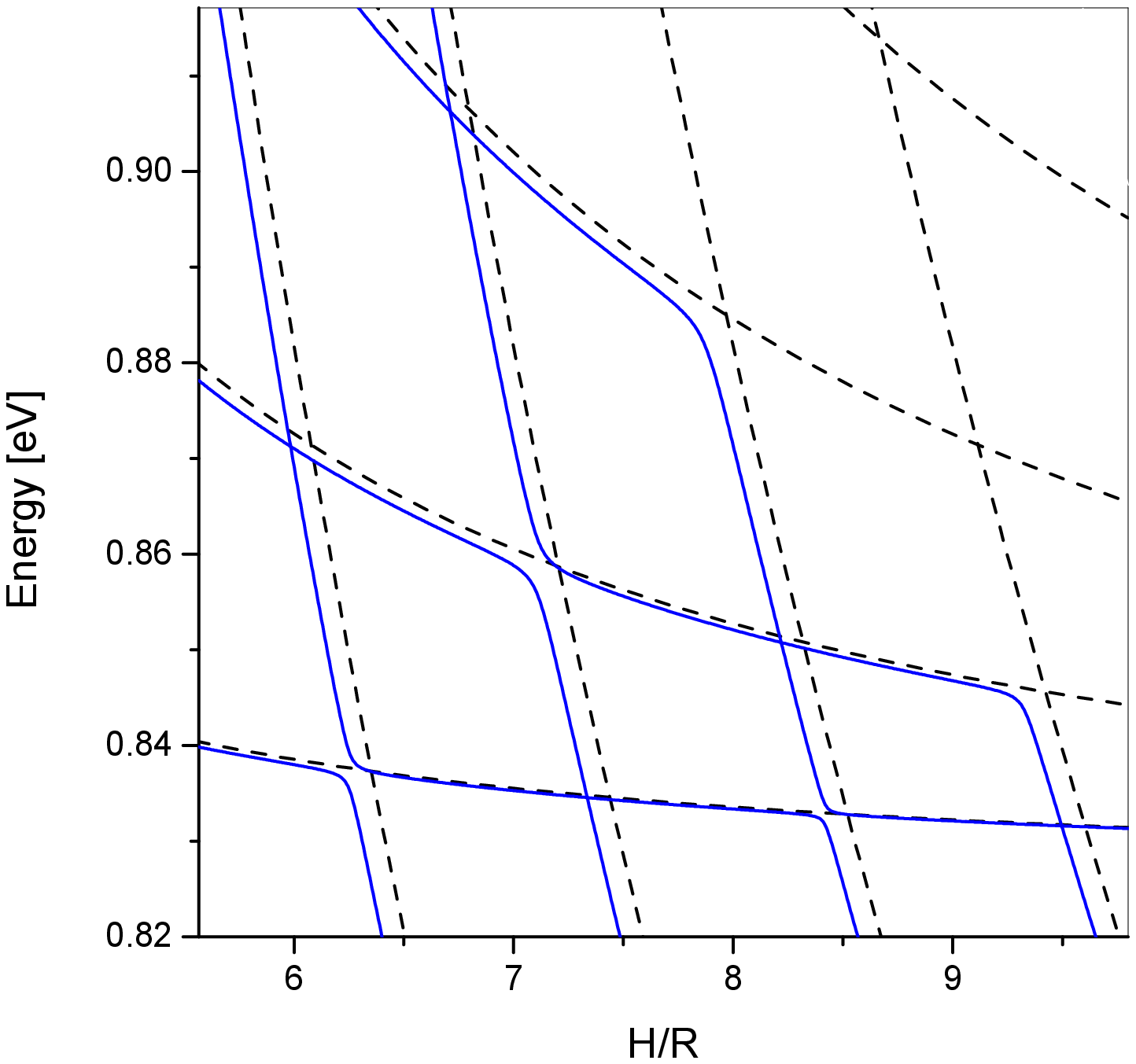}
\caption{(Color online) Same as Fig.~1 but on an
enlarged scale.}
\end{figure}

The fact that the solid lines in Fig.~1 closely follow the dashed ones suggests
that the quantum number $n_z$ and the wave function~(\ref{frod}) represent a good
zero-order approximation to describe the motion along the nanorod growth direction,
$z$. On the other hand,
one can see that the dependencies of nanorod energy levels presented in Fig.~1 demonstrate both crossings
and anti-crossings (or avoided crossings). In order to reveal the
underlying pattern some regions of Fig.~1 are presented in Figs.~2 and~3 on an
enlarged scale. These figures reveal that the level crossings and anti-crossings
form a checker-board pattern, although the anti-crossings become less pronounced
as the nanorod height increases.  The same situation occurs for other values of 
$M$, as demonstrated in Fig.~4 for $M=3/2$. In Fig.~5 the levels for $|M|=1/2$ and
$|M|=3/2$ are plotted together to account for lowest conduction band energy levels
in nanorods.
\begin{figure}[ht]
  \centering
    \includegraphics[width=.4\textwidth]{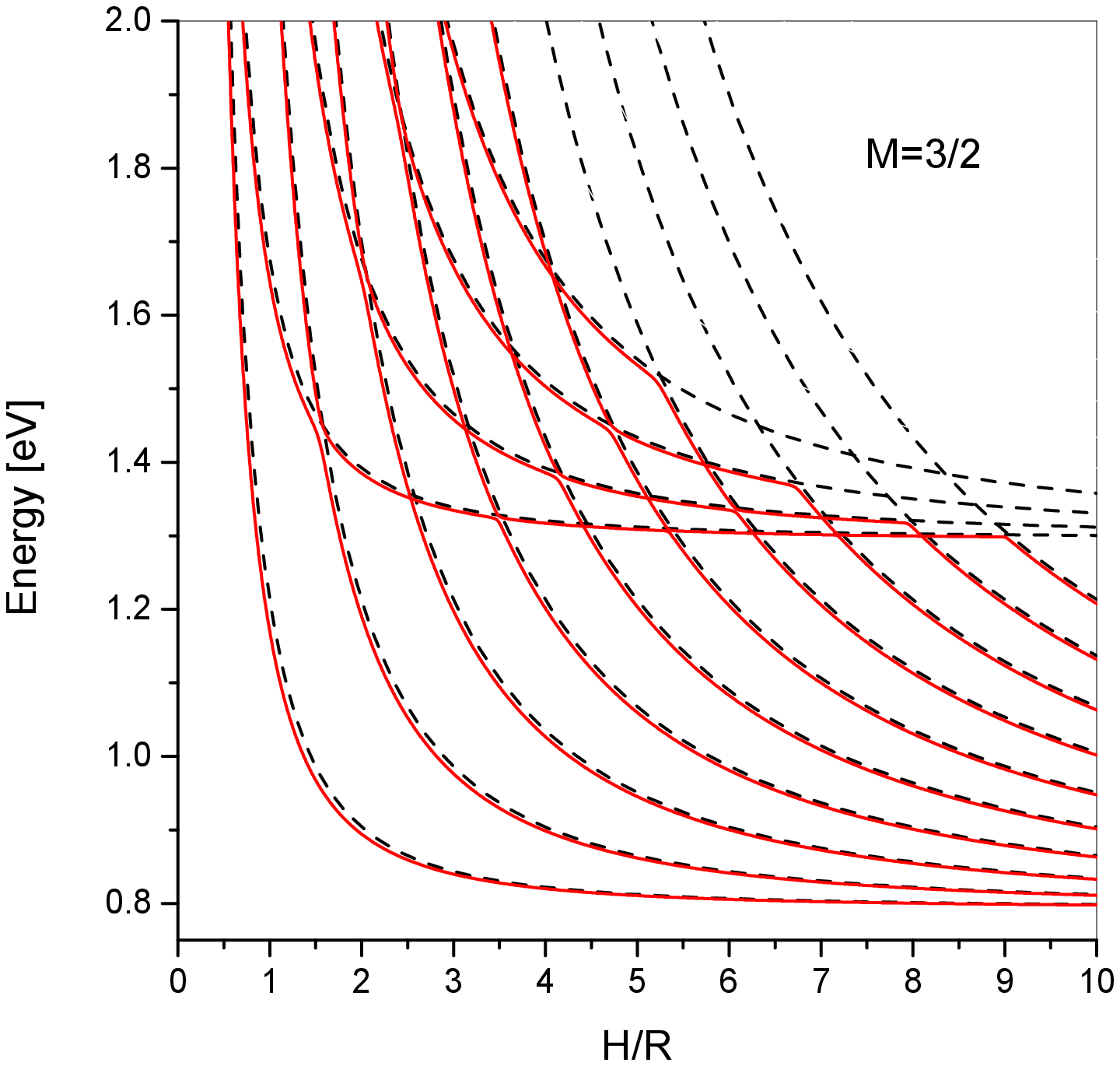}
\caption{(Color online) Same as Fig.~1 but for $M=3/2$.}
\end{figure}
\begin{figure}[htb]
  \centering
    \includegraphics[width=.4\textwidth]{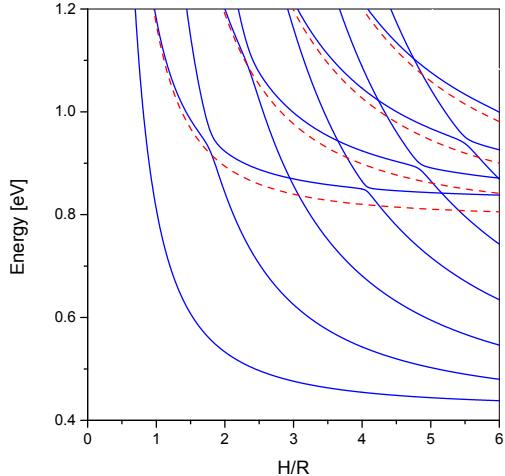}
\caption{(Color online) Energies of levels of size quantization for the 
quantum number $|M|=1/2$ (solid lines) and $|M|=3/2$ (dashed lines) in a PbSe
nanorod of radius $R=20$~\AA\mbox{} and height $H$ as functions of $H/R$. }
\end{figure}

The occurrence of the checker-board pattern of crossings and anti-crossings
in the height dependence of electron energy levels in lead salts
nanorods can be understood if one analyzes the equation describing
electron energy dispersion in lead salts quantum wires~\cite{mycom}. This 
equation is given by Eq.~(25) of Ref.~\cite{mycom}. For a given value of the quantum number $M$ it has the following structure
\begin{equation}
\label{main}
k_z^2 \, A +B \, C=0 \,.
\end{equation}
When $k_z=0$ Eq.~(\ref{main}) breaks into two separate equations $B=0$ and
$C=0$ describing quantum states of opposite parities 
at the apexes of 1D subbands~\cite{mycom}. When $k_z \neq 0$
the two solutions mix. This mixing is fairly weak and can be considered as
a perturbation of the subband dispersion relations described by the equations 
$B=0$ and $C=0$. The structure of Eq.~(\ref{main}) suggests that this perturbation is linear in $k_z$. In the space of envelope wave functions $k_z$ can be considered
as the operator of the $z$ component of the linear momentum in the momentum representation. The same operator in the coordinate representation is proportional to the first derivative with respect to $z$, or $\partial /\partial z$. The functions~(\ref{frod}) are either odd or even with respect to the transformation $z \rightarrow H-z$. 
The 
operator $\partial /\partial z$ can only have non-vanishing matrix elements between the functions
of different parity (under the transformation $z \rightarrow H-z$). 
Thus, the levels originating from the lower quantum wire subband characterized in the zeroth order by odd (even) $n_z$ cross with the levels originating from the upper quantum wire subband characterized by odd (even) $n_z$
and anti-cross with the levels characterized by even (odd) $n_z$, provided that these subbands have the same quantum number $|M|$. Note, however, that, if the states in the first ($n_c=1$) quantum wire subband with a given $|M|$ are characterized at $k_z=0$
by the equation $B=0$ so will the states from the third ($n_c=3$) subband. Therefore, the electron energy levels in nanorods originating from the first and the third 
quantum wire subbands characterized by the same quantum number $|M|$ will always cross, and the checker board pattern will be broken for higher energies. This 
observation has been verified by numerical calculations.

To summarize, we have demonstrated that for most practical purposes the effect of the finite height of a lead salt nanorod on its electronic structure can be accounted for by sampling the subband dispersion dependencies of the infinite cylindrical quantum wire of the same radius at wave numbers $k_z=\pi n_z/H$, where
$H$ is the nanorod height and $n_z$ is an integer. This approximation can fail
if, for a given nanorod height, some of the energy levels originating from different
quantum wire subbands characterized by the same absolute value $|M|$ of the total angular momentum projection onto the quantum wire axis are close to one another. The levels
would repel if the corresponding states are characterized by the numbers
$n_z$ of different parities and originate from subbands characterized by the same value of $|M|$ and main quantum numbers of opposite parities, giving rise to the levels splitting.

This work  
was supported by the Research Corporation for Science Advancement
(Award No.~20081), the Russian Foundation for Basic Research and the National Science Foundation (Grant No. HRD-0833178).

\end{document}